\documentclass[superscriptaddress,twocolumn,pre]{revtex4}
\usepackage{amsmath}
\usepackage{amssymb}
\usepackage{graphicx}
\usepackage{hyperref}
\usepackage{amsfonts}
\usepackage{bbm}
\usepackage{comment}
\usepackage{color}
\usepackage[usenames,dvipsnames,svgnames,table]{xcolor}

\hypersetup{colorlinks,citecolor=RoyalBlue,filecolor=black,linkcolor=Mulberry,urlcolor=PineGreen}

\begin{document}
\title{Point Defects, Chirality and Singularity Theory in Cholesteric Liquid Crystal Droplets} 
\author{Joseph Pollard}
\affiliation{Mathematics Institute, Zeeman Building, University of Warwick, Coventry, CV4 7AL, United Kingdom.}
\author{Gregor Posnjak}
\affiliation{Condensed Matter Department, Jo\v{z}ef Stefan Institute, Jamova 39, 1000 Ljubljana, Slovenia.}
\author{Gareth P. Alexander}
\affiliation{Department of Physics and Centre for Complexity Science, University of Warwick, Coventry, CV4 7AL, United Kingdom.}
\author{Simon \v{C}opar}
\affiliation{Department of Physics, Faculty of Mathematics and Physics, University of Ljubljana, Jadranska 19, 1000 Ljubljana, Slovenia.}
\author{Igor Mu\v{s}evi\v{c}}
\affiliation{Condensed Matter Department, Jo\v{z}ef Stefan Institute, Jamova 39, 1000 Ljubljana, Slovenia.}
\affiliation{Department of Physics, Faculty of Mathematics and Physics, University of Ljubljana, Jadranska 19, 1000 Ljubljana, Slovenia.}

\begin{abstract}
We develop a theory of point defects in cholesterics and textures in spherical droplets with normal anchoring. The local structure of chiral defects is described by singularity theory and a smectic-like gradient field establishing a nexus between cholesterics and smectics mediated by their defects. We identify the defects of degree $-2$ and $-3$ observed experimentally with the singularities $D_4^{-}$ and $T_{4,4,4}$, respectively. Radial point defects typical of nematics cannot be perturbed into chiral structures with a single handedness by general topological considerations. For the same reasons, the spherical surface frustrates the chirality in a surface boundary layer containing regions of both handedness.
\end{abstract}
\date{\today}
\maketitle 

\section*{Introduction}

The character of materials is conveyed by their defects: They control strength or fragility~\cite{taylor1934}, determine elastic interactions~\cite{poulin1997,skarabot2007}, mediate self-assembly~\cite{musevic2006,Musevic,wang2016}, and precipitate phase transitions~\cite{abrikosov1957,berezinskii1971,kosterlitz1973,renn1988,kikuchi2007}. The widespread influence of defects derives from their high energetic cost, strong elastic distortions, and topological nature. This is especially true in the liquid crystalline mesophases, whose defects have often offered key insights and motivated the naming of textures. Moreover, liquid crystals offer a versatile setting for studying topological phenomena, often with relevance across multiple disciplines, including cosmological strings~\cite{chuang1991}, biological tissues and morphogenesis~\cite{saw2017,bouligand2008}, and magnetic Skyrmion textures~\cite{nych2017,ackerman2017,machon2016}. Chiral materials furnish an especially rich source of topological and geometric phenomena, and chirality is also at the heart of one of the fundamental contrasts given in liquid crystals; between smectics and cholesterics. In de Gennes' famous analogy~\cite{deGennes1972}, their contrast is equivalent to the difference between metals and superconductors, with the expulsion of twist from smectics being analogous to the Meissner effect. Here, we will emphasise a new paradigm where the defects in chiral materials are forced to have the local structure of a smectic in order to be chiral.

The character of point defects in cholesteric liquid crystals (or magnetic Skyrmion textures) appears not to have been considered previously, apart from recent experimental work realising them~\cite{posnjak2016,posnjak2017,GregorThesis}. This contrasts with the situation in nematics where point defects, known colloquially as hedgehogs, have been extensively studied over several decades~\cite{poulin1997,skarabot2007,musevic2006,nabarro1972,kurik1982,volovik1983,pargellis1991,kleman2006,Lavrentovich2001}. They can be generated deliberately as satellite defects to colloidal inclusions, or in droplets, with normal anchoring. For spherical colloids, point defects form elastic dipoles and fascilitate self-assembly of colloidal chains and lattices~\cite{Musevic}, while in droplets transitions between defect states produced by changes in boundary conditions~\cite{volovik1983} can provide highly sensitive sensors~\cite{lin2011,lee2016}. Point defects in nematics are classified by an integer ($\pi_2(\mathbb{RP}^2)\cong\mathbb{Z}$) known as the hedgehog charge, or degree~\cite{alexander2012}. Normal anchoring boundary conditions on a surface of genus $g$ correspond to a degree $1-g$ and induce compensating point defects of the same total degree in liquid crystal surrounding colloidal inclusions~\cite{senyuk2013} or inside handlebody droplets~\cite{pairam2013}. In addition to their own phenomenology the topological character of point defects has also provided fundamental insight into disclination loops and their classification~\cite{alexander2012,copar2011PRL,copar2011PRE}. The lack of a similar body of work for point defects in cholesterics represents a gap in our understanding of chiral materials; closing it will provide considerable insight into cholesterics and Skyrmion textures of chiral ferromagnets~\cite{milde2013}.

In recent experiments~\cite{posnjak2016,posnjak2017,GregorThesis} point defects were created in spherical droplets of cholesteric liquid crystal with normal surface anchoring and shown to have markedly different properties from their nematic counterparts. Whereas spherical nematic droplets relax to a state of minimal distortion with a single defect of degree $+1$ near the centre and an approximately radial texture~\cite{kleman2006}, a diverse variety of different states were found in the cholesteric case. These included point defects of degree $+1, -1, -2$ and $-3$ -- the higher charge defects observed for the first time -- as well as numerous string-like `constellations' and `topological molecules'. Defects were located both in the centre of the droplet and in close proximity to its surface. The latter are not the boojums associated to planar anchoring~\cite{volovik1983} but a feature of the cholesteric order.

Here, we provide a theoretical description of point defects in cholesterics that reproduces the structures observed experimentally. The main feature of this description is that the structure is determined by a gradient field, more commonly associated with smectics, whose basic properties include the fundamental absence, or expulsion, of chirality~\cite{renn1988,deGennes1972}. This establishes a previously unnoticed nexus between cholesterics and smectics. Generically, point defects are not chiral and sit on surfaces separating domains of opposite handedness; there is then an energetic drive to expel those defects from the interior. For those that are chiral, their local structure is that of a gradient field with isolated critical point and the chirality provided by higher order terms in a Taylor series. The type of critical point provides a label for the defect that refines the topological degree. Generic chiral point defects are described by Morse critical points with Morse index 1 or 2 (only). Defects with higher topological charge are associated to degenerate critical points, giving a new physical application for singularity theory~\cite{Thom1973,berry1976,PostonStewart}; we identify those of charge $-2$ and $-3$ seen experimentally with the $D_4^{-}$ and $T_{4,4,4}$ singularities, respectively. Annihilations of defects, their splittings and metamorphoses are described by unfoldings of the singularities. Line defects in the pitch axis of the cholesteric ($\lambda$ lines) are also prescribed by the gradient field of the singularity. Finally, the spherical surface of the droplet frustrates the chirality in a topological way that necessitates the existence of localised regions near the surface with the wrong handedness.
The similarity between cholesterics and chiral ferromagnets means that this description also applies to the monopoles that have been seen in Skyrmion states~\cite{milde2013}.

\section*{Cholesterics}

Liquid crystals are anisotropic fluids described by an average molecular orientation called the director field, ${\bf n}$, a unit magnitude vector, although the symmetry of the nematic phase is that of a line field, {\sl i.e.} ${\bf n}\sim -{\bf n}$. The director field extremises the Frank free energy~\cite{frank1958} 
\begin{equation}
F = \int \frac{K_1}{2} \bigl( \nabla \cdot {\bf n} \bigr)^2 + \frac{K_2}{2} \bigl( {\bf n} \cdot \nabla \times {\bf n} + q_0 \bigr)^2 + \frac{K_3}{2} \bigl( ( {\bf n} \cdot \nabla ) {\bf n} \bigr)^2 \, dV ,
\label{eq:Frank}
\end{equation}
where the $K_i$ are elastic constants and $q_0$ is a chiral coupling constant, called the chirality. We shall assume that the cholesteric is right-handed ($q_0>0$). In a bulk material, the minimum energy configuration is the cholesteric ground state ${\bf n}=\cos q_0z \,{\bf e}_x + \sin q_0z \, {\bf e}_y$. It is periodic with periodicity $p=\pi/q_0$, known as the cholesteric half-pitch. A dimensionless measure of the strength of the chirality in the droplet is the ratio of its diameter $2R$ to the half-pitch, $N=2R/p$. In the experiments of Refs.~\cite{posnjak2017,GregorThesis}, typical values were $2\lesssim N\lesssim 6$; in the simulations shown here we use values between 2 and 4. We consider a spherical droplet with normal anchoring boundary conditions that has only point defects in the interior. The director field can then be oriented and we orient it to coincide with the outward radial direction, ${\bf n}={\bf x}/\|{\bf x}\|$, on the surface. This choice imposes that the degrees of all of the interior defects add up to $+1$. Numerical minimisation of Eq.~\eqref{eq:Frank} is done using a finite difference relaxation algorithm on a cubic grid; for simplicity we adopt a one elastic constant approximation ($K_1=K_2=K_3$) and the only relevant parameter is then the dimensionless ratio $N$. This captures all the phenomenology and major experimental observations; more quantitative comparison can come from including defect core structure and elastic anisotropy. 


\section*{Achiral Point Defects}

\begin{figure*}[t]
\centering
\includegraphics[width=\textwidth]{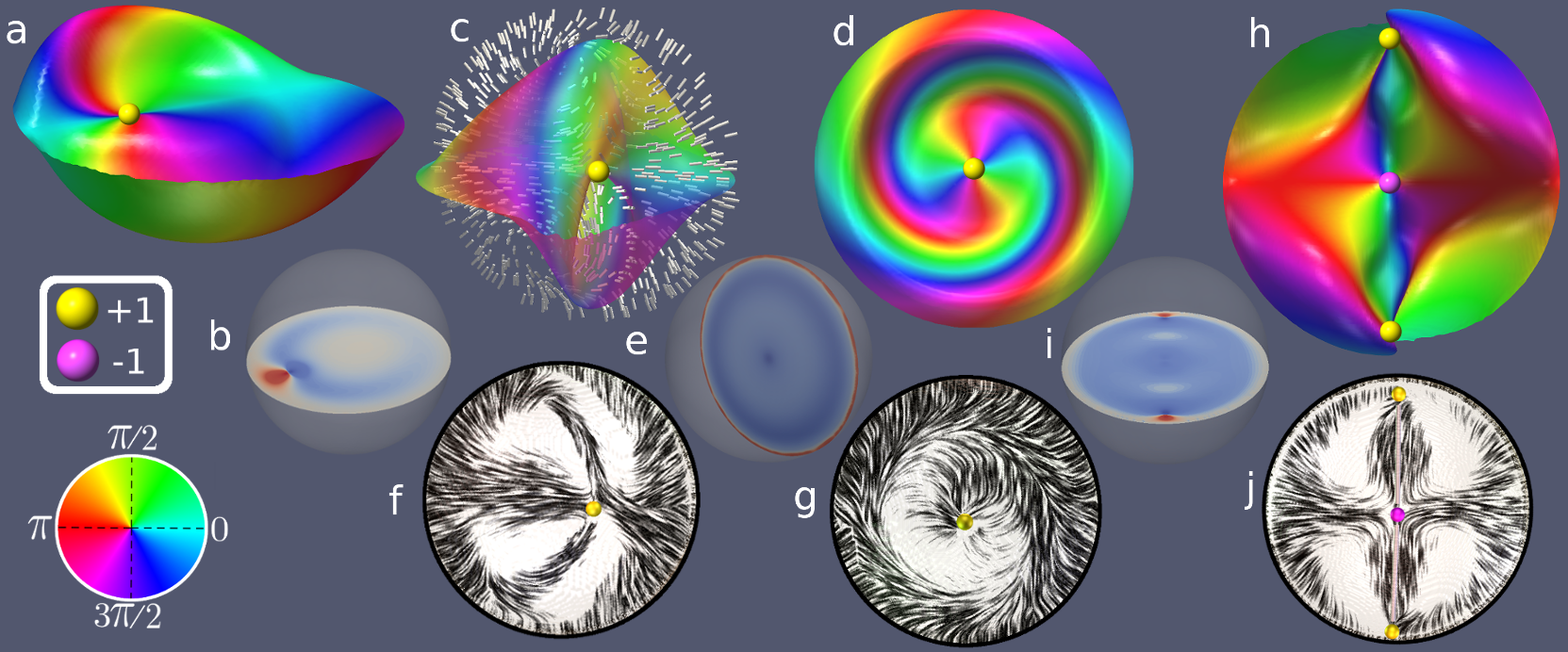}
\caption{(a) Generic, achiral defect with degree $+1$. The surface is the locus $n_z=0$ (Pontryagin-Thom surface), with colour indicating the angle between the $x$ and $y$ components of the director. (b) The twist on a midplane $xy$-slice showing the reversal of twist at the location of the defect; left-handed regions are shown in red, and right-handed regions in blue. (c) $n_z =0$ and (d) $n_x = 0$ surfaces for a chiral defect with Morse index 2. In (c) the director has also been shown on a midplane $xz$-slice for added clarity. In (d) the viewing direction is along $x$. (e) The twist on a $yz$-slice containg the defect. Note that the region around the defect is uniformly right-handed, but that there is a region of reversed twist near the surface. Experimental observations of this defect are shown in (f) and (g). (h) Chiral defect with Morse index 1 and degree $-1$; viewing direction along $z$. (i) The twist on a midplane $xy$-slice containing all three defects; again, note that the central defect is uniformly chiral. (j) An experimental observation of the same defect. In all panels, defects have been highlighted and colour-coded according to degree.}
\label{fig:morse}
\end{figure*}

We represent the director field as ${\bf n}={\bf m}/\|{\bf m}\|$ where ${\bf m}$ is a vector field with isolated zeros at each of the point defects. Since ${\bf n}\cdot\nabla\times{\bf n}=({\bf m}\cdot\nabla\times{\bf m})/\|{\bf m}\|^2$ the director field is chiral whenever ${\bf m}$ is. We shall call a defect (zero in ${\bf m}$) chiral if it is surrounded by a region in which the vector field is twisted of a single handedness; otherwise we shall call it achiral. The twist is frustrated at the defects -- if ${\bf m}={\bf 0}$ then evidently ${\bf m}\cdot\nabla\times{\bf m}=0$ -- so that it is non-generic for a defect to be chiral. To explain the frustration, at the level of a structural classification the zeros in ${\bf m}$ are locally equivalent to a polynomial form and in the generic case to a linear vector field. If the curl is non-zero then the twist will be a linear function and the zero is not chiral. An example is the vector field ${\bf m}=x \,{\bf e}_x - z \,{\bf e}_y + y \,{\bf e}_z$ for which the twist is ${\bf m}\cdot\nabla\times{\bf m}=2x$. This structural form provides a good local description of the defects that arise close to the droplet surface in experiments.

We show in Fig.~\ref{fig:morse}(a) the result of a typical simulation of a weakly chiral droplet ($N<1$). The texture is visualised through a Pontryagin-Thom surface~\cite{chen2013}, which shows the set of points where the director is horizontal ($n_z=0$) and coloured according to the horizontal orientation. The most noticeable change from the radial profile of a nematic droplet is the displacement of the defect from the centre towards the surface. This can be understood by looking at the twist throughout the droplet, which we show in Fig.~\ref{fig:morse}(b); the defect sits on a surface that separates the droplet into regions with opposite handedness. There is then an energetic drive to expand the region with the correct handedness (blue) at the expense of that with the wrong sense of twist (red) and the balance of this with the increased elastic distortion determines the position of the point defect. The broken spherical symmetry from the displacement of the defect defines an axis through the droplet and the chiral distortion of the director field turns this into the axis of a double twist cylinder.

Increasing $N$ for small values continues this trend; the defect moves further towards the surface to reduce the size of the region with the wrong handedness and the axis it defines becomes increasingly recognisable as a double twist cylinder. However, this does not continue forever and for values of $N \gtrsim 2$ the defect starts to move away from the surface again as the director field restructures continuously to the form shown in Fig.~\ref{fig:morse}(c), where the point defect has become chiral; the twist is shown in Fig.~\ref{fig:morse}(e) and is uniformly right-handed (blue) around the defect. We describe this structure now. 

\section*{Chiral Point Defects}

Generic linear vector fields fail to be chiral because their curl is non-zero. Consequently, the local form for a chiral zero is ${\bf m} = \nabla \phi + {\bf m}_c$, where $\phi$ has an isolated critical point at the origin and ${\bf m}_c$ is of higher order in a Taylor series. As alluded to, the structure of the director field is then unexpectedly close to that of a smectic, normal to the level sets of $\phi$, or smectic layers. The type of critical point in $\phi$ is a label for the chiral zero that refines the degree. In the generic case, the critical point is described by a quadratic function, referred to as Morse-type. Quadratic functions can be diagonalised and up to equivalence are distinguished by the number of negative terms in this diagonal form, called the Morse index~\cite{Milnor}.

The generic chiral defect of degree $+1$ and Morse index 2 is described by the local form
\begin{equation}
{\bf m} = ax \,{\bf e}_x - (y-qxz) \,{\bf e}_y - (z+qxy) \,{\bf e}_z ,
\label{eq:Morse2}
\end{equation}
for which the twist is ${\bf m}\cdot\nabla\times{\bf m}=-q(2ax^2+y^2+z^2)$. Here $a>0$ is dimensionless and $q$ has dimensions of an inverse length. For the particular case $a=2$, ${\bf n}\cdot\nabla\times{\bf n} \approx -q$ suggesting that we should take $q=q_0$ up to an $O(1)$ factor. Changing the sign of the vector field \eqref{eq:Morse2} gives the local form of a generic chiral defect of degree $-1$ and Morse index 1. Figure~\ref{fig:morse}(c), (d) show the results of simulations initialised using \eqref{eq:Morse2}, or its negative (h), throughout the central portion of the droplet and a radial profile outside. The correspondence with experimental observations~\cite{posnjak2016,posnjak2017,GregorThesis} is striking, Fig.~\ref{fig:morse}(f), (g) and (j).

The twist of the director field on a slice through the droplet is shown in Fig.~\ref{fig:morse}(e) and (i). Throughout the interior, including at the defect, the twist is right-handed (blue) but close to the surface there are regions where it is left-handed (red). This behaviour is topological, as we describe presently.

The radial hedgehog ${\bf m} = x \,{\bf e}_x + y \,{\bf e}_y + z \, {\bf e}_z$ corresponds to the gradient field of the function $\phi = \tfrac{1}{2}(x^2+y^2+z^2)$, with Morse index 0, and is the observed configuration for nematic droplets. However, this structure is not preserved in a chiral droplet and exhibits fundamental frustration with a state of uniform twist; there is no choice of higher order term ${\bf m}_c$ that makes the radial hedgehog chiral. This follows from a theorem of Eliashberg \& Thurston~\cite{Eliashberg}, the Reeb stability theorem for confoliations, which we describe informally. It is based on the property that the level sets of $\phi$ are spheres. Consider any such level set and separate it into two hemispheres $D_1\cup D_2$ with common boundary an equator. The twist over each disc can be related to the nature of the director field on its boundary. This is based on the properties of integral curves of the plane field orthogonal to ${\bf n}$, {\sl i.e.} curves whose tangent is always perpendicular to the director. When the twist vanishes, the director is the normal to the level sets of $\phi$ and any curve lying in such a surface is integral. This applies in particular to the boundary of any disc. When the twist is non-zero the director is no longer perpendicular to the boundary. In this case, the boundary can be lifted -- pushed up or down along the surface normal direction -- to create an integral curve. This curve is not closed; its endpoints have a vertical displacement between them whose magnitude is related to the non-vanishing of the twist~\cite{Eliashberg}. Importantly, if the twist is right-handed then the displacement is along the positive normal direction, while if it is left-handed the displacement will be in the negative direction. Applying this to the separation of the spherical level set into two hemispheres, the displacement around the boundary of each hemisphere is equal and opposite, since the two boundaries are the same equator but traversed in opposite directions. As a result, if the twist is right-handed in one hemisphere it will be left-handed in the other, and in equal measure. We see from this that radial hedgehogs are incompatible with the preferred handedness of cholesterics. 

\section*{Droplet Surface and Boundary Layer}

Precisely the same argument applies to a boundary layer at the droplet surface where the director becomes radial to match the boundary conditions. By the same theorem of Eliashberg \& Thurston the twist cannot be uniformly right-handed and by necessity there is some region of reversed twist close to the surface. In the case of the Morse index 2 defect (Fig.~\ref{fig:morse}(c)) this takes the form of a non-singular ring, while for Morse index 1 (Fig.~\ref{fig:morse}(h)), and all other examples, the regions of twist reversal are associated to achiral $+1$ `surface defects' and this topological argument provides an explanation for the presence of such defects localised close to the surface. This boundary behaviour is peculiar to spherical surfaces and there is no corresponding topological limitation on the twist (handedness) for surfaces of any positive genus ($g>0$), whose boundary behaviour is characterised by a one-dimensional cohomology class of the surface, {\sl i.e.} an element of $\mathbb{R}^{2g}$~\cite{Eliashberg}. 

Regions of opposite twist localised near point defects close to surfaces with normal anchoring have recently been observed in experiments and numerical simulations~\cite{ackerman2016}. In those experiments the surface is the flat plane of a glass slide, as opposed to being spherical. It is only spherical surfaces that have a topological requirement for regions of reversed twist. 

\section*{Singularity Theory and Defects of Higher Degree}

Morse critical points have degree $\pm 1$ and so cannot describe the defects with higher degree ($-2$, $-3$) observed experimentally~\cite{posnjak2017}; they correspond to degenerate critical points, for which we follow Arnold's classification~\cite{Arnold1998}. 

\begin{figure*}[t]
\centering
\includegraphics[width=\textwidth]{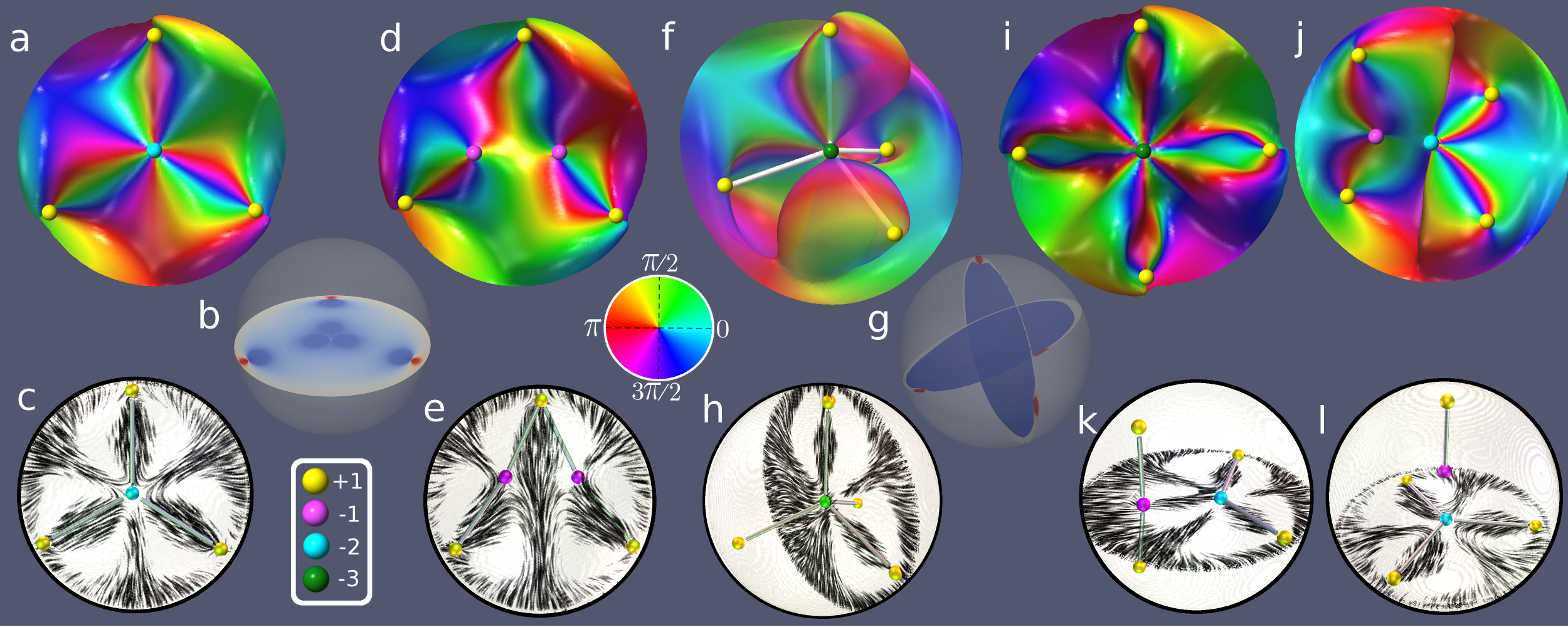}
\caption{(a) Point defect with degree $-2$ associated to the $D_4^{-}$ singularity. The surface is the locus $n_z=0$ (Pontryagin-Thom surface), with colour indicating the angle between the $x$ and $y$ components of the director. (b) The twist on a midplane $xy$-slice containing all four defects. (c) An experimental realisation of the degree $-2$ defect. (d) An unfolding of $D_{4}^{-}$ produces a `V-shaped constellation' closely similar to (e) the experimental observation. (f) Point defect with degree $-3$ associated to the $T_{4,4,4}$ singularity (`bonds' between defects added for visual clarity), along with (g) the twist on perpendicular slices containing the defects and (h) an experimental realisation. (i) The $X_9$ singularity has lower codimension than $T_{4,4,4}$ and also has degree $-3$ but not the tetrahedral symmetry observed in experiments. (j) Unfoldings of $T_{4,4,4}$ produce the `topological molecules' observed experimentally, as shown in (k). (l) A different unfolding of $T_{4,4,4}$ that was also observed in experiments. In all panels, defects have been highlighted and colour-coded according to degree.}
\label{fig:D4}
\end{figure*}

Among the simple singularities, the only models for defects with degree $-2$ are furnished by the classes $D_{2k}^{-}$. The simplest such model is $D_4^-$, which we write as
\begin{equation}
\phi = \kappa \biggl( x^2 y - \frac{1}{3} y^3 \biggr) + \frac{1}{2} z^2 ,
\label{eq:D4}
\end{equation}
where $\kappa$ is a constant with dimensions of an inverse length. The gradient vector field can be made chiral in a neighbourhood of the origin by adding the germ
\begin{equation}
{\bf m}_c = q \bigl[ - \kappa z \bigl( x^2-y^2 \bigr) {\bf e}_x + 2 \kappa xyz \,{\bf e}_y \bigr] - q^2 z^3 \bigl[ - y \,{\bf e}_x + x \,{\bf e}_y \bigr] .
\label{eq:D4chiral}
\end{equation}
The correspondence between this local model and experimental observations~\cite{posnjak2017,GregorThesis} is striking (Fig.~\ref{fig:D4}(a), (c)), although we find in our simplified director-based simulations that the defect is unstable and breaks apart into two degree $-1$ defects, each with the Morse-type local structure described above. This points to the importance of defect core structure, or elastic anisotropy, for stability in physical systems. 

The experimentally observed defect with degree $-3$~\cite{posnjak2017,GregorThesis} (Fig.~\ref{fig:D4}(h)) is reproduced by the unimodal singularity $T_{4,4,4}$, which we write
\begin{equation}
\phi = a xyz + \frac{1}{4} \bigl( x^4 + y^4 + z^4 \bigr) ,
\label{eq:T444}
\end{equation}
where the modulus $a$ has dimensions of length. This is a codimension 9 singularity with multiplicity 11, characteristics which convey its high complexity. The codimension of a singularity is the number of independent parameters, or perturbations, that resolve it into simpler pieces, while the multiplicity is the number of Morse critical points that it splits into (as a complex polynomial) under a generic perturbation.

In addition to the origin, where there is an isolated zero of degree $-3$, the gradient vector field has isolated zeros of degree $+1$ (and Morse index 0) at the four points $a(1,-1,1), a(-1,1,1), a(1,1,-1), a(-1,-1,-1)$, corresponding to the vertices of a tetrahedron. This suggests that the modulus $a$ should take a value of $R/\sqrt{3}$, up to an $O(1)$ factor, so that these defects sit near the surface of the droplet, as observed experimentally. The gradient field of \eqref{eq:T444} can be perturbed into a chiral point defect via a generic method that we describe in the following section. Simulations initialised with the $T_{4,4,4}$ singularity produce a numerically stable degree $-3$ point defect surrounded by four tetrahedrally-arranged surface defects in excellent agreement with the experiment (Fig.~\ref{fig:D4}(f), (h)).

The appearance of singularity theory in the classification of the zeros of chiral vector fields represents an interesting new physical application for this branch of mathematics~\cite{Thom1973,berry1976,PostonStewart,Arnold1998} and it is natural to consider how the formalism of singularity theory manifests itself in this setting. In particular, the unfoldings of singularities provide models for the combination and splitting of chiral point defects. The annihilation of two chiral defects with opposite degree is given generically by the unfolding of the $A_2$ singularity and the germ
\begin{equation}
\phi = \frac{\kappa}{3} x^3 + \frac{1}{2} y^2 - \frac{1}{2} z^2 + c x ,
\label{eq:A2}
\end{equation}
where $\kappa$ is a constant with dimensions of an inverse length and $c$ is a parameter of the unfolding with dimensions of length. When $c/\kappa$ is positive there are no critical points, while when it is negative there are isolated Morse critical points at $(\pm |c/\kappa|^{1/2},0,0)$ with Morse indices 1 ($+$) and 2 ($-$).

The unfolding of the $D_4^{-}$ singularity provides a generic description of the splitting of a degree $-2$ chiral defect, or merging of two point defects with the same degree ($-1$)  
\begin{equation}
\phi = \kappa \biggl( x^2 y - \frac{1}{3} y^3 \biggr) + \frac{1}{2} z^2 + c_1 x + c_2 y + \frac{c_3}{2} \bigl( x^2+y^2 \bigr) .
\label{eq:D4unfolding}
\end{equation}
As an example of the unfolding, when only $c_1$ is non-zero the $D_4^{-}$ singularity splits into two Morse-type defects, at the points $(\pm |c_1/2\kappa|^{1/2},\pm |c_1/2\kappa|^{1/2},0)$ when $c_1/\kappa$ is negative and at $(\mp |c_1/2\kappa|^{1/2},\pm |c_1/2\kappa|^{1/2},0)$ when $c_1/\kappa$ is positive. The resulting pattern of point defects within the droplet is shown in Fig.~\ref{fig:D4}(d) and is strongly reminiscent of the `V-shaped constellations' observed experimentally (Fig.~\ref{fig:D4}(e))~\cite{posnjak2017,GregorThesis}. Other configurations of `topological molecules' observed experimentally~\cite{posnjak2017,GregorThesis} are consistent with unfoldings of higher degree germs; examples for degree $-3$ are shown in Fig.~\ref{fig:D4}(j-l). 

At the same time, not all aspects of singularity theory have an immediate realisation in cholesterics. For instance, it is not clear whether the full list of singularities all occur, or in an order predicted by their codimension. As an example, the lowest codimension singularity with degree $-3$ is $X_9$ (codimension 7) and not $T_{4,4,4}$. In Fig.~\ref{fig:D4}(i) we show the structure of a cholesteric droplet generated by the $X_9$ germ; its fourfold symmetry with the surface defects sitting in a plane and at the vertices of a square immediately distinguishes it from the tetrahedral $T_{4,4,4}$. In the latter, the surface defects are maximally separated, which lowers the elastic free energy. 

\section*{Curl Eigenfields and Harmonic Critical Points}

A curl eigenfield, also known as a Beltrami field, is a vector ${\bf v}$ for which $\nabla \times {\bf v} = \lambda {\bf v}$ for a scalar eigenvalue $\lambda$~\cite{chandrasekhar1957,etnyre2000}. Setting ${\bf n}={\bf v}/\|{\bf v}\|$ we obtain ${\bf n}\cdot\nabla\times{\bf n}=\lambda$ so that a normalised curl eigenfield has constant twist away from its zeros. This suggests a special model, that corresponds to minimisers of the Frank free energy~\eqref{eq:Frank} when $K_2 \gg K_1, K_3$. In this limit, the free energy density is minimised pointwise except at the zeros of the curl eigenfield, giving a construction reminiscent of that for the blue phases~\cite{wright1989} or their helimagnet analogues~\cite{tewari2006,binz2006}.

Curl eigenfields are divergenceless and in the local form ${\bf m} = \nabla \phi + {\bf m}_c$, $\phi$ is a harmonic function. The chiral perturbation is also determined by $\phi$; as a formal series, ${\bf m}_c=\sum_{j} {\bf m}_c^{(j)}$, we have the recursive equations
\begin{align}
& \nabla \times {\bf m}_c^{(1)} = - q_0 \nabla \phi , && \nabla \times {\bf m}_c^{(j)} = - q_0 {\bf m}_c^{(j-1)} ,
\end{align}
together with $\nabla \cdot {\bf m}_c^{(j)}=0$. Each term, ${\bf m}_c^{(j)}$, is determined up to the addition of a harmonic gradient. Thus the entire structure of such a chiral point defect is determined by the gradient field of a harmonic function. Many of the singularities in Arnold's list have harmonic representatives; for instance the Morse singularity of~\eqref{eq:Morse2} is harmonic if $a=2$, and a harmonic germ for the $T_{4,4,4}$ singularity is
\begin{equation}
\phi = a xyz + \frac{1}{4} \bigl( x^4 + y^4 + z^4 - 3 x^2 y^2 - 3 y^2 z^2 - 3 z^2 x^2 \bigr) .
\label{eq:harmonicT444}
\end{equation}
They can all be perturbed so as to be chiral following the prescription for the germs of curl eigenfields. Nonetheless, it is well-known that not all types of critical point can occur in a harmonic function: By the maximum principle, there are no local maxima or minima. This recapitulates the content of Eliashberg \& Thurston's more general result~\cite{Eliashberg}; there are no chiral point defects with radial director field.

Germs of harmonic critical points can be expressed in terms of spherical harmonics, $\phi = \sum_{l,m}a_{l m} r^l Y_{l m}$. For example, the harmonic Morse critical points are given by $r^2 Y_{2 0}$, while the harmonic form of $T_{4,4,4}$, equation \eqref{eq:harmonicT444}, is $\sqrt{8\pi/105}\,a \,\textrm{Im}\, r^3 Y_{3 2} + \sqrt{5\pi/126}\, \textrm{Re}\, r^4 Y_{4 4} + \sqrt{\pi}/6\, r^4 Y_{4 0}$. Similarly, a harmonic germ for the degree $-2$ defect can be constructed in this way, using $\phi = \textrm{Im} \, r^3 Y_{3 3} + \kappa r^4 Y_{4 0}$. The chiral perturbation is also given explicitly as ${\bf m}_c^{(1)}=q_0 \sum_{l,m} \frac{a_{l m}}{l+1} r^l \, {\bf x} \times \nabla Y_{l m}$. In general, the leading spherical harmonic determines the symmetry of the defect and, in particular, the arrangement of accompanying `surface defects', while higher order terms ensure that it is isolated. In this way, defects with higher degrees and desired symmetries can be constructed. The observed point defects of degrees $-1, -2$ and $-3$ have symmetries such that the accompanying $+1$ `surface defects' exhibit highly geometric configurations with maximal separation; antipodal points, the vertices of an equilateral triangle, and the vertices of a tetrahedron, respectively. These correspond to the energy minimising configurations of identical point charges on the surface of a sphere, interacting via the Coulomb force, known as the Thomson problem~\cite{thomson1904}, which provides a candidate for the symmetry of the higher degree defects. A degree $-5$ defect with octahedral symmetry is captured by the germ $\phi = - \sqrt{7/10}\, r^4 Y_{4 0} - \textrm{Re} \,r^4 Y_{4 4}$. Octahedral arrangements of `surface defects' have been observed experimentally in `topological molecules' consistent with unfoldings of the degree $-5$ germ~\cite{posnjak2017,GregorThesis}. Curiously, the bipyrmidal degree $-4$ defect requires higher order spherical harmonics to construct; it is captured by the germ $\phi = \textrm{Im}\, r^5 Y_{5 3} + \tilde{\kappa}\, \textrm{Re} \, r^8 Y_{8 0}$. Its higher order structure suggests a less stable core, which may explain why it has not yet been observed. 

\section*{$\lambda$ Lines and Umbilics: Defects in the Cholesteric Pitch}

\begin{figure*}[t]
\centering
  \includegraphics[width=1.0\linewidth]{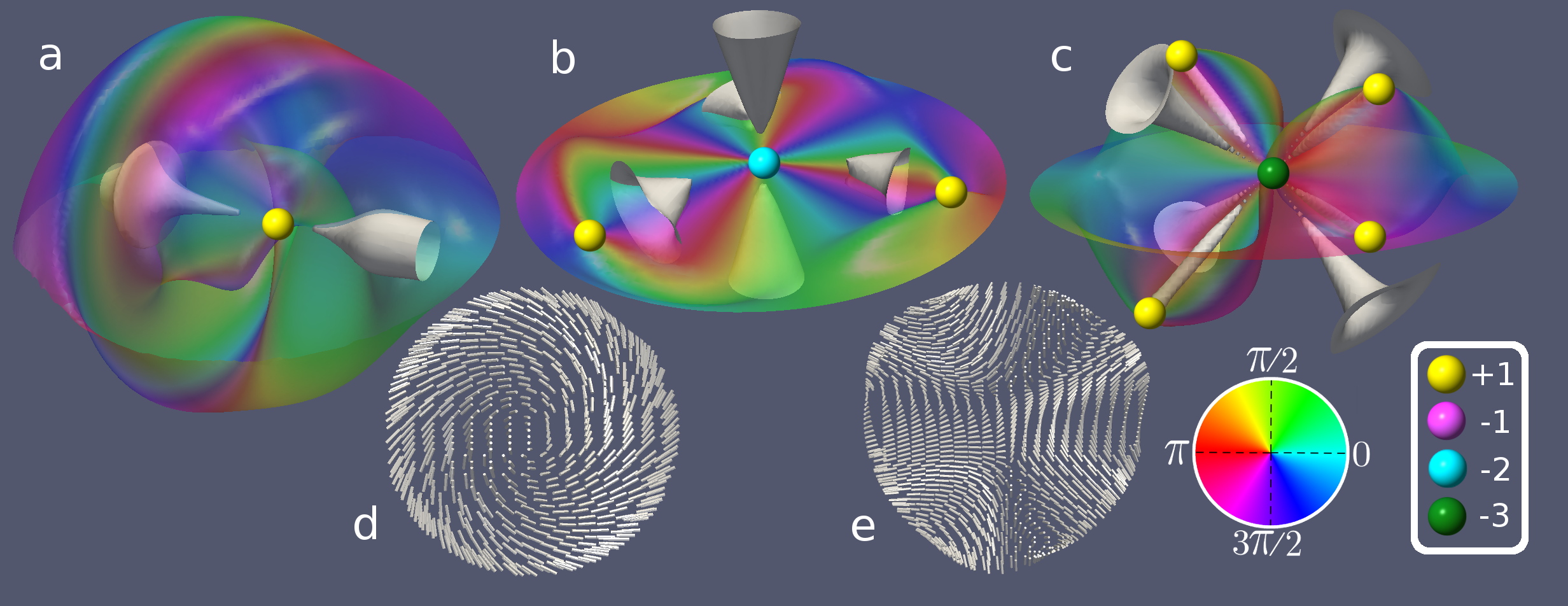}
  \caption{Umbilic lines for the (a) Morse index 2, (b) $D^-_4$ and (c) $T_{4,4,4}$ singularities, indicated by white isosurfaces. Each defect has double twist cylinders connecting it to the degree $+1$ `surface defects'. The remaining umbilics are $\lambda^{-1/2}$-lines that terminate on the droplet surface. Typical cross-sectional profiles for the umbilics are shown in (d) double twist cylinders and (e) $\lambda^{-1/2}$-lines. In all panels, defects have been highlighted and colour-coded according to degree.}
  \label{fig:umbilics}
\end{figure*}

In a cholesteric, the local direction along which the director twists is called the pitch axis. It varies through the material and can have its own defects, called $\lambda$ lines~\cite{Lavrentovich2001,beller2014}. Point defects in the director are associated with a confluence of $\lambda$ lines, whose number and type are related on topological grounds to the type of point defect~\cite{machon2016}; for a defect of degree $k$ there are a total of $4|k|$ lines, counted with multiplicity.

In fact, a much more detailed description can be given: As the structure of chiral point defects is determined by the gradient field of an isolated singularity, the location and type of the defects in the pitch axis coincides with the structure of umbilic lines of the local level manifolds of the function germ. Passing through the critical value the level sets change between $1+|k|$ disconnected discs and a connected surface of Euler characteristic $1-|k|$. Each disc is pierced by the axis of a double twist cylinder ($\lambda$/umbilic line of multiplicity $2$), while a further $2|k|-2$ elementary lines pierce the connected surface along directions of high symmetry. The pattern of lines for the Morse index 2, $D_4^{-}$ and $T_{4,4,4}$ singularities is shown in Fig.~\ref{fig:umbilics}, with the umbilic lines indicated by white tubes. Taking $D_4^{-}$ as an example, the degree is $-2$ so that there must be $8$ umbilic lines, counted with multiplicity. There are three degenerate umbilic lines (double twist cylinders -- Fig.~\ref{fig:umbilics}(d)), connecting the central defect to the $+1$ surface defects. These should be counted with multiplicty $2$~\cite{machon2016}, leaving a deficit count of $2$. Symmetry considerations tell us that this is made up by two additional umbilic lines ($\lambda^{-1/2}$ profile -- Fig.~\ref{fig:umbilics}(e)) along the $z$-axis. A similar description can be given in all cases. 

\section*{Discussion}


Smectics and cholesterics are often contrasted~\cite{beller2014}; sometimes this emphasises similarlties, for instance they have the same elasticity~\cite{toner1981}, but usually the focus is on the sense in which they are antithetical. The non-vanishing of the twist is precisely the condition -- Frobenius integrability theorem -- that the director field is not the normal to any set of layers and provides a geometric characterisation of cholesterics~\cite{machon2017}. Conversely, the expulsion of twist from a smectic is a hallmark of that phase, famously analogous to the Meissner effect~\cite{deGennes1972}. But in the presence of defects the division is diminished: Chiral point defects are determined by the gradient field of an isolated singularity, yielding a convergence between cholesterics and smectics. 

Defects are central to understanding the behavior and properties of materials. Their link with singularity theory suggests that the richness of that subject can be harnessed to direct future developments. More generally, the topological frustration of chirality by radial defects and spherical surfaces gives new fundamental insight into the character of chiral materials, applicable to Skyrmions in chiral ferromagnets as well as cholesterics. 

\acknowledgments{GPA acknowledges beneficial discussions with Michael Berry and David Chillingworth on singularity theory and Randy Kamien on defects in the pitch axis. This work was supported by the UK EPSRC through Grant No.~EP/L015374/1 (JP \& GPA) and the Slovenian Research Agency (ARRS) under contracts P1-0099 (GP, S\v{C} \& IM) and J1-9149 (S\v{C}).}

\end{document}